\documentclass[12pt]{iopart}
\usepackage{iopams}
\begin{document}
\letter{Interpreting Stone's model of Berry phases}
\author{Paolo Carra} 
\address{ European Synchrotron Radiation Facility,
B.P. 220, F-38043 Grenoble C\'{e}dex, France}
\ead{carra@esrf.fr} 
\date{\today} 
\begin{abstract}
We show that a simple quantum-mechanical model,  
put forward by Stone sometime ago, affords
description of site magnetoelectricity, a phenomenon which takes
place in crystals (and molecular systems) when space inversion is
locally broken and coexistence of electric and magnetic moments
is permitted by the site point group. We demonstrate this by 
identifying a local order parameter, which is odd under both space 
inversion and time reversal. This order parameter (a magnetic 
quadrupole) characterises Stone's ground state. Our results indicate that the
model, extended to a lattice of sites, could be relevant 
to the study of electronic properties of transition-metal oxides.
A generalisation of Stone's hamiltonian to cover cases of different 
symmetry is also discussed.
\end{abstract}
\submitto{\JPA}
\pacs{71.27+a}
\maketitle
In 1986 Stone reported a study of the Hamiltonian 
\begin{equation}
H=\case{1}{2I}{\bi L}^2 -\mu\, {\bi n}\cdot{\bsigma}\, ,
\label{hamiltonian}
\end{equation}
which he performed by path-integral and by conventional 
quantum-mechanical techniques (Stone 1986). 
The conventional method considers the basis set of spinor spherical
harmonics   
\begin{equation}
|j\pm\case12,jm\rangle=\sum_{m',\xi}
C_{j\pm\case12,m';\case12,\xi}^{j,m}
|j\pm\case12,m'\rangle|\case12,\xi\rangle \, ,
\label{spinor_sph_har}
\end{equation}
as suggested by $[H,{\bi J}]_-=0$, with ${\bi J}=
{\bi L} +\frac{1}{2}{\bsigma}$; ${\bi L}$ stands for the angular 
momentum operator that generates rotations of 
${\bi n}={\bf r}/r$; ${\bsigma}$ are Pauli matrices and 
$I$ is the rotor moment of inertia; $[,]_-$denotes a commutator and
$C_{j_1,m_1;j_2,m_2}^{j,m}$ is a Clebsch-Gordan coefficient. 
On account of the 
property (Stone 1986, Varshalovich \etal 1988) 
\begin{equation}
{\bi n}\cdot{\bsigma}|j\pm\case{1}{2},jm\rangle
= - |j\mp\case{1}{2},jm\rangle\,,
\label{scalar}
\end{equation}
the Hamiltonian $H$ is readily diagonalised in this basis.
In the large-$\mu$ limit, the ground state is found to be
\begin{equation}
|g\rangle_-= \case{1}{\sqrt{2}}\left(|j+\case12,jm\rangle - 
|j-\case12,jm\rangle\right)\,,
\label{ground_state}
\end{equation}
with energy $E=-\mu$.

Stone's motivation was to provide a simple quantum-mechanical 
example in which the Berry phase gives rise to Wess-Zumino terms. 
Indeed, for large $\mu$, Eq. (\ref{hamiltonian}) describes the 
motion of a constrained spin, which is equivalent to motion 
of a charged particle about a magnetic monopole (Leinaas 1978). 

Equation (\ref{hamiltonian}) has been interpreted as a model for a solenoid, 
which is rotating about its centre of mass where a spin-$\case12$ particle 
is placed. When $\mu$ is small, the solenoid and the particle would 
spin independently. As $\mu$ becomes large, the spin will become 
slaved to the direction of the solenoid (Stone 1986, Aitchinson 1987). 
Notice that this physical picture leads to the
coupling ${\bi B}\cdot{\bsigma}$, which is space and time
even. (${\bi B}$ stands for the magnetic field generated by the solenoid.) 
Such a symmetry should be contrasted with that of 
${\bi n}\cdot{\bsigma}$, which is space and time odd.  

The current Letter will demonstrate that the model defined 
by Eq. (\ref{hamiltonian}) affords description of a different effect: 
{\it site magnetoelectricity}. Such a phenomenon occurs in crystalline
and molecular systems when space-inversion symmetry is locally broken and
co-existence of electric and magnetic moments is permitted by the pertinent 
site point group. An effective magnetoelectric interaction between
these two moments would be described by Eq. (\ref{hamiltonian}) provided
we identify ${\bi n}$ with a unitary electric-dipole moment. 
(The electric charge 
$e$ is merged into $\mu$.) This new interpretation of the model
does not affect its dynamical regimes (Stone 1986), which remain 
those of the rotating 
solenoid with ${\bi B}$ replaced by ${\bi n}$.  

As shown by Goulon and his collaborators (Goulon \etal 2000, 2002),
microscopic magnetoelectric behaviour of crystals can be investigated 
using near-edge absorption of x rays, which implies excitations of 
inner-shell electrons to empty valence states. As is known, this experimental 
technique is site selective, a feature resulting from the
tuning of x-ray energy at a given inner-shell threshold. 
Sensitivity to the long-range order of 
local magnetoelectric order parameters is obtained by recording dichroic 
signals which stem from an interference between electric-dipole and 
electric-quadrupole transitions. 
As a consequence, scalars (e.g. ${\bi n}\cdot{\bsigma}$) are not probed 
by these experiments, which detect the long-range order of local (on-site)
magnetoelectric order parameters represented by one-particle irreducible tensors of
rank 1,2 and 3. 
One set of these order parameters specifically serve
our purposes: the {\it magnetic quadrupoles} 
(rank-2 tensors) \footnote{A simple mechanical model of a magnetic 
quadrupole is provided by two parallel coils run through by opposite 
currents.}. In the LS-coupling scheme, they read
\begin{eqnarray}
&&
{\mathcal M}^{(2)}_L \equiv [{\bi n},{\bi L}]^{(2)}\,,
\quad {\mathcal M}^{(2)}_S \equiv [{\bi n},{\bi S}]^{(2)}\,,
\label{ls_magneticquadrupoles}
\\
&&
{\mathcal M}^{(2)}_T\equiv\case{\sqrt{3}}{\sqrt{2}}
[i[{\bOmega}_L,{\bi L}]^{(2)}, {\bi S}]^{(2)}\,,\quad
{\mathcal M}^{(2)}_F \equiv \case{\sqrt{35}}{2}
[[{\bi n}, {\mathcal Q}^{(2)}]^{(3)}, {\bi S}]^{(2)}\,,
\nonumber
\end{eqnarray}
as shown by recent theoretical work on x-ray dichroism and resonant 
scattering in noncentrosymmetric crystals 
(Carra \etal 2003, Marri and Carra 2004).
(The symbol $[\,,]^{(k)}$ denotes Clebsch-Gordan coupling of irreducible 
tensors; the spin operator is defined by ${\bi S} = \frac{1}{2}{\bsigma}$; 
 ${\mathcal Q}^{(2)}=[{\bi L},{\bi L}]^{(2)}$ denotes an orbital quadrupole.) 
Inspection of magnetic nature identifies orbital, spin and 
spin-orbital elements in the set. 
All tensors in (\ref{ls_magneticquadrupoles}) 
are space and time odd, being thereby invariant under the combined 
action of these transformations. 

A set of vector order parameters will also be considered in connection
with Stone's model. Its elements are defined by (Marri and Carra, 2004)
\begin{equation}
{\bi n},\quad {\bi P}_S\equiv{\bOmega}_L\times{\bi S} \quad 
{\rm and}\quad {\bi P}_T \equiv-\case{2\sqrt{5}}{\sqrt{3}}
[[{\bi n},{\bi L}]^{(2)}, {\bi S}]^{(1)} \,,
\label{ls_polarisation} 
\end{equation}
in LS coupling. These irreducible tensors have {\it polar} (electric)
symmetry, i.e., they are space odd and time even 
\footnote{The LS tensors 
(\ref{ls_magneticquadrupoles}-\ref{ls_polarisation}) are given in compact 
forms. In this representation irrational prefactors appear. They can
be removed by recoupling transformations. We find 
$
[{\bi S}, [{\bi L},{\bi n}]^{(2)}]^{(1)}= -\case{\sqrt{3}}{2\sqrt{5}}
\left({\bi S}\cdot{\bi L}\,{\bi n} + {\bi L}\,{\bi S}\cdot{\bi n}
\right),
$
and similarly for
$
[{\bi S}, [{\bOmega}_L,{\bi L}]^{(2)}]^{(1)};
$
furthermore, 
$
[ {\bi S}, -i[{\bOmega}_L,{\bi L}]^{(2)}]^{(2)}
= \case{\sqrt{2}}{\sqrt{3}}\left([{\bi S}\times{\bOmega}_L, {\bi L}]^{(2)}
+\case{1}{2}[{\bi S}, {\bOmega}_L\times{\bi L}]^{(2)}\right)
$
and 
$
[{\bi S}, [{\bi n}, {\mathcal Q}^{(2)}]^{(3)}]^{(2)}
=\left(3[{\bi S},[{\bi n},{\mathcal Q}^{(2)}]^{(2)}]^{(2)}
+6[[{\bi S},{\mathcal Q}^{(2)}]^{(2)},{\bi n}]^{(2)}\right.
-\left. 3[{\bi S}\cdot{\bi n},{\mathcal Q}^{(2)}]^{(2)}
-5[{\bi S},[{\mathcal Q}^{(2)},{\bi n}]^{(2)}]^{(2)}\right)/\sqrt{35}\,.
$
}.

The order parameters (\ref{ls_magneticquadrupoles}-\ref{ls_polarisation}) 
are defined in the second quantisation formalism. For example,
\begin{equation}
({\mathcal M}^{(2)}_L)_q 
= \sum_{l,l'=l\pm1\atop m,m',\sigma,\sigma'}
\case{1}{2}\left[
\langle \sigma'|\langle l'm'|
({\mathcal M}^{(2)}_L)_q |lm\rangle|\sigma\rangle 
c^{\dagger}_{l'm'\sigma'}
c^{\phantom{\dagger}}_{lm\sigma}
+{\rm h.c.}\right]\,,
\label{second_quant} 
\end{equation}
and similarly for the others. Here, $c^{\dagger}_{lm\sigma}$ and
$c^{\phantom{\dagger}}_{lm\sigma}$ denote fermionic creation and 
annihilation operators, respectively.

As stated, the main purpose of the current work is to discuss 
magnetoelectric properties of Stone's Hamiltonian in the large 
$\mu$ limit. To this end, we will show that the symmetry 
property displayed by the scalar ${\bi n}\cdot{\bsigma}$, 
when acting on the spinor spherical harmonics 
$|j\pm\case{1}{2},jm\rangle$ [Eq. (\ref{scalar})], 
extends to the irreducible tensors defined by expressions 
(\ref{ls_magneticquadrupoles}-\ref{ls_polarisation}), leading to 
relations whereby the magnetoelectric behaviour 
of $|g\rangle_-$ is readily inferred.
 
The basis set $|j\pm\case{1}{2},jm\rangle$ provides
a convenient framework for describing {\it parity-breaking 
electron hybridisation} (e.g. $pd$ mixing in transition-metal oxides), 
in the jj coupling scheme \footnote{As shown below, in this basis, 
parity breaking hybribisation displays rotational symmetry [$SU(n)$, 
in the general case]. Rather than by ordinary $|jm\rangle$, 
irreducible representations are spanned by 
$\case{1}{\sqrt{2}}(|j+\case{1}{2},jm\rangle \pm |j-\case{1}{2},jm\rangle)$ 
for ranks 0 and 2, and by $\case{1}{\sqrt{2}}(e^{i\pi/4}|j+
\case{1}{2},jm\rangle \pm 
e^{-i\pi/4}|j-\case{1}{2},jm\rangle)$ for rank 1.}.  
It is thus clear that to generalise Eq. (\ref{scalar}) to higher-rank 
order parameters we must first determine the 
form of the corresponding tensors in jj coupling.
This is accomplished by resorting to the theory of LS$\rightarrow$jj 
transformations (Edmonds, 1974). In the case of space-odd irreducible 
tensors, such transformations contain matrix elements between 
the states $|jm\rangle$ and $|j'm'\rangle$,
with $j=l\pm\frac{1}{2}$ and $j'=l'\pm\frac{1}{2}$. Solving the 
corresponding equations for $j'=j$, as demanded by $[H,{\bi J}]_-=0$, 
provides the required jj-coupled order parameters, which will appear 
as linear combinations of space-odd LS-coupled irreducibe tensors.
(Notice that ${\bi n}\cdot{\bsigma}/2={\bi n}\cdot{\bi J}$, as 
${\bi n}\cdot{\bi L}=0$; no transformation is needed in this case.) 
Technically, the derivation is laborious as it implements
angular-momentum recoupling methods (Racah calculus). Thus, for 
convenience of the reader, we will first state our results, 
then illustrate their physical content and, at the end, discuss 
mathematical aspects of the formulation. 

The required jj-coupled magnetic quadrupole is found to be
\begin{eqnarray}
\widetilde{\mathcal M}_J^{(2)}(l',l) &=&\case{1}{5}
\left(\case{l+l'-1}{2}\right)\left(\case{l+l'+3}{2}\right)
{\mathcal M}^{(2)}_S(l',l)+\case{2}{3}{\mathcal M}^{(2)}_T(l',l)
\nonumber
\\
&+&\case{1}{5}{\mathcal M}^{(2)}_F(l',l)-\case{1}{2}
{\mathcal M}^{(2)}_L(l',l)
\label{jj_magneticquadrupole} 
\end{eqnarray}
yielding, after some algebra, 
\begin{equation}
\sum_{l,l'=l\pm1}
\widetilde{\mathcal M}_J^{(2)}(l',l)_z|j\pm\case{1}{2},jm\rangle
 = -\case{3m^2-j(j+1)}{\sqrt{6}}|j\mp\case{1}{2},jm\rangle\,,
\label{rank_2}
\end{equation}
with $\hat{\bf z}$ the quantisation axis.
This result generalises Eq. (\ref{scalar}) and shows that 
$|g\rangle_-$ is an eigenstate of the jj-coupled magnetic 
quadrupole operator; in other words, the large-$\mu$ ground state 
of Stone's hamiltonian is magnetoelectric. 
Our conclusion is further supported by what follows. 
Consider the jj-coupled operator
\begin{equation}
\widetilde{\bi P}_J(l',l)= {\bi n}(l',l)+{\bi P}_S(l',l)
-2{\bi P}_T(l',l)\,.
\label{jj_polarisation}
\end{equation}
We have
\begin{equation}
\sum_{l,l'=l\pm1}
\widetilde{\bi P}_J(l',l)_0|j\pm\case{1}{2},jm\rangle 
= -m|j\mp\case{1}{2},jm\rangle\,,
\end{equation}
showing that $|g\rangle_-$ is an eigenstate of the jj-coupled
unitary electric-dipole moment, with eigenvalue
$m$. $|g\rangle_-$ is thus characterised by the simultaneous presence 
of an electric and a magnetic moment in a parallel (as expected) 
configuration. [Reversing the sign of the coupling constant in Stone's model
($\mu\rightarrow-\mu$, large $\mu$) would change the ground state 
to $|g\rangle_+= \case{1}{\sqrt{2}}\left(|j+\case12,jm\rangle + 
|j-\case12,jm\rangle\right)$, which is characterised by an 
antiparallel alignement of the moments and by a 
magnetic quadrupole with opposite sign.]   

Thus, we are led to the
conclusion that, in the strong coupling limit, the magnetoelectric 
interaction can be viewed as a problem of a 
magnetic moment constrained by an electric field. 
In turn, this is equivalent to the problem 
of a charged particle moving in the field of a magnetic 
monopole, both classically and quantum mechanically (Leinaas 1978). 
This seems to tally with recent work on magnetic monopoles 
in crystal momentum space (Fang \etal 2003).

According to our findings, Stone's model provides a good
starting point in the study of interactions between (local) 
electric and magnetic moments in crystals.
For this purpose, an extension of the model to a lattice of sites is
now needed. Such a model, characterised 
by an order parameter which
violates space inversion and time reversal, could be relevant 
in the analysis of electronic properties of transition-metal oxides.

In certain site point groups (Cracknell 1975), magnetoelectric
interactions result in configurations where electric
and magnetic moments are mutually perpendicular. Such interactions arise 
from toroidal distributions of currents and are described by anapole 
moments. 
In LS coupling, the set of anapolar order parameters reads 
(Carra \etal 2003, Marri and Carra 2004)
\begin{eqnarray}
&&
{\bOmega}_L\equiv\case12({\bi n}\times{\bi L}-
{\bi L}\times{\bi n})\,,\quad
{\bOmega}_S\equiv{\bi n}\times {\bi S}\,, 
\nonumber
\\
&&{\bOmega}_T\equiv-\case{2\sqrt{5}}{\sqrt{3}}
[[{\bOmega}_L,{\bi L}]^{(2)}, {\bi S}]^{(1)}\,.  
\label{ls_anapoles}
\end{eqnarray}
The required jj-coupled anapole takes the form
\begin{equation}
\widetilde{\bOmega}_J(l',l)= 
\case{2}{l+l'+1}{\bOmega}_L(l',l)-(l+l'+1){\bOmega}_S(l',l) 
 + \case{4}{l+l'+1}{\bOmega}_T(l',l)\,,
\label{jj_anapole}
\end{equation}
giving
\begin{equation}
(\widetilde{\bOmega}_J)_z |j\pm\case{1}{2},jm\rangle
=\sum_{l,l'=l\pm1}
\widetilde{\bOmega}_J(l',l)_z|j\pm\case{1}{2},jm\rangle
= \mp im |j\mp\case{1}{2},jm\rangle\,.
\end{equation}
$|\tilde{g}\rangle_-=\case{1}{\sqrt{2}}
(e^{i\pi/4}|j+\case{1}{2},jm\rangle - 
e^{-i\pi/4}|j-\case{1}{2},jm\rangle)$ 
is therefore an eigenstate of the jj-coupled anapole 
operator. It is readily shown that 
$|\tilde{g}\rangle_-$ is the large-$\mu$ limit ground state 
of the hamiltonian obtained by replacing ${\bi n}\cdot{\bsigma}$ with
$\hat{\bf z}\cdot\widetilde{\bOmega}_J$ in Eq. (\ref{hamiltonian}). 
Notice that $|\tilde{g}\rangle_-$ is also an eigenstate 
of a jj-coupled {\it pseudodeviator}, a polar rank-2 tensor, 
with eigenvalue $[3m^2-j(j+1)]/\sqrt{6}$. This new hamiltonian thus 
displays x-ray natural circular dichroism. [The 
derivation of this result will not be given for lack of space. LS 
pseudodeviators have been discussed by Marri and Carra (2004).]

We conclude with some remarks concerning technical aspects of 
our derivation.
Generalising the concept of coupled double tensor 
(Judd, 1967), we define
\begin{equation}
w^{(xy)z}_{\zeta}(l',l)=\sum_{\xi,\eta,\lambda,\lambda',\sigma,\sigma'}
C^{z\zeta}_{x\xi;y\eta}C^{y\eta}_{\frac{1}{2}\sigma';\frac{1}{2}\sigma}
C^{x\xi}_{l'\lambda';l\lambda}c^{\dagger}_{l'\lambda'\sigma'}
\tilde{c}^{\phantom{\dagger}}_{l\lambda\sigma}+\,\,{\rm h.c.}
\label{judd_ls1}
\end{equation}
and
\begin{equation}
v^{(j'j)z}_{\zeta}(l',l) = \sum_{m,m'}C^{z\zeta}_{j'm';jm}
c^{\dagger}_{l',j'm'}\tilde{c}^{\phantom{\dagger}}_{l,jm}
+\,\,{\rm h.c.}\,,
\end{equation}
where $\tilde{c}^{\phantom{\dagger}}_{l\lambda\sigma}=
(-1)^{l-\lambda+\frac{1}{2}-\sigma} 
c^{\phantom{\dagger}}_{l-\lambda-\sigma}$ and 
$\tilde{c}^{\phantom{\dagger}}_{l,jm}=(-1)^{j-m}
{c}^{\phantom{\dagger}}_{l,j-m}$, so that creation and annihilation
operators transform as the components of irreducible tensors.
[Notice that $w^{(x0)x}(l',l)$ are spinless, whereas 
$w^{(x1)z}(l',l)$ depend on spin.] 
The importance of $w^{(xy)z}(l',l)$  
lies in the fact that all one-electron LS order parameters defined 
by (\ref{ls_magneticquadrupoles}-\ref{ls_polarisation}) can be 
expressed as multiples of them (Wigner-Eckart theorem). For example,
\begin{equation}
{w}^{(20)2}(l',l)= 
-\case{\sqrt{2}}{\sqrt{l(l+1)}C_{l0;10}^{l'0}
\left\{ {1\,\,1\,\,2\atop l'\,\,l\,\,l}
\right\}}
{\mathcal M}^{(2)}_L (l',l)\,.
\end{equation}
In a similar way, one-electron jj order parameters can be expressed
as multiples of $v^{(j'j)z}(l',l)$.
The tensors $w^{(xy)z}(l',l)$ and $v^{(j'j)z}(l',l)$ are 
related by a standard LS$\rightarrow$jj transformation (Edmonds, 1974), 
which reads
\begin{equation}
w^{(xy)z}(l',l)=\sum_{j,j'}(-1)^{x+y+z}[x,y,j,j']^{\frac{1}{2}}
\left\{ 
\begin{array}{ccc}
l & l' & x \\
\frac{1}{2} & \frac{1}{2}  & y \\
j & j' &z
\end{array} 
\right\}
v^{(j'j)z}(l',l) \,,
\label{jj_transformation}
\end{equation}
with 
$[a,...,b]=(2a=1)\cdot\cdot\cdot(2b+1)$. In the
case of magnetic-quadrupole order parameters, Eq. (\ref{jj_transformation}) 
yields a system of four equations. By solving this system 
for $j'=j$ and $l'=l\pm1$, we find
\begin{eqnarray*}
&&\widetilde{\mathcal M}_J^{(2)}(l',l) 
= {\rm r.h.s\;of\; Eq.\;(\ref{jj_magneticquadrupole})}
\\ 
&& = -\case{3}{2}(2l+1)(2l'+1)
\left\{
[{\bi n},{\bi J}]^{(l'+\frac{1}{2},l-\frac{1}{2})2}\,\delta_{l',l-1}
+[{\bi n},{\bi J}]^{(l'-\frac{1}{2},l+\frac{1}{2})2}\,\delta_{l',l+1}
\right\}\,.
\end{eqnarray*}

Vector order parameters lead to systems of three equations. 
Their solutions for $j'=j$ and $l'=l\pm1$ read
\begin{equation}
\widetilde{\bi P}_J(l',l) = 
-\case{3(l+l'+1)}{2}\left[
{\bi n}_J^{l'+\frac{1}{2},l-\frac{1}{2}}\,\delta_{l',l-1}
+{\bi n}_J^{l'-\frac{1}{2},l+\frac{1}{2}}\,\delta_{l',l+1}
\right]
\end{equation}
and
\begin{equation}
\widetilde{\bOmega}_J(l',l)= 
-\case{4(2l+1)(2l'+1)}{l+l'+1}\left[
{\bOmega}_J^{l'+\frac{1}{2},l-\frac{1}{2}}\,\delta_{l',l-1}
+{\bOmega}_J^{l'-\frac{1}{2},l+\frac{1}{2}}\,\delta_{l',l+1}
\right]\,,
\label{detail}
\end{equation}
providing a full definition of Eqs. (\ref{jj_polarisation}) and 
(\ref{jj_anapole}). In Eq. (\ref{detail}), 
\begin{eqnarray}
{\bOmega}_J^{j'j}(l',l)&=&\sum_{m,m'}\langle l',j'm'|
({\bi \nabla}_{\Omega}-{\bi \nabla}_{\Omega}^{\dagger})\times{\bi J}
+{\bi J}\times({\bi \nabla}_{\Omega}-{\bi \nabla}_{\Omega}^{\dagger})
|l, jm\rangle
\nonumber 
\\
&&
c^{\dagger}_{l',j'm'}\tilde{c}^{\phantom{\dagger}}_{l,jm}\,,
\label{jjx_anapole}
\end{eqnarray}
where ${\bi \nabla}_{\Omega}=-i{\bi n}\times{\bi L}$
\footnote{A natural choice for the anapole in jj 
coupling would
be $({\bi n}\times{\bi J}-{\bi J}\times {\bi n})/2$ 
(Dothan and Ne'eman, 1966). However, this operator cannot be 
employed in our case
as its matrix element vanishes for $j'=j$. The vector appearing 
in Eq. (\ref{jj_anapole}) removes this drawback (Jerez, 2003).}. 

In the case of anapolar order parameters, 
LS Judd's tensors are defined
as 
\begin{equation}
\overline{w}^{(xy)z}(l',l)=i\sum_{\xi,\eta,\lambda,\lambda',\sigma,\sigma'}
C^{z\zeta}_{x\xi;y\eta}C^{y\eta}_{\frac{1}{2}\sigma';\frac{1}{2}\sigma}
C^{x\xi}_{l'\lambda';l\lambda}c^{\dagger}_{l'\lambda'\sigma'}
\tilde{c}^{\phantom{\dagger}}_{l\lambda\sigma}+\,\,{\rm h.c.}\,
\label{judd_ls2}
\end{equation}
and similarly in jj coupling.\\ \\
I am grateful to M. Fabrizio, A. Jerez, E. Katz and T. Ziman for discussions 
and a critical reading of the manuscript.
\section*{References} 
\begin{harvard}
%
\item[] Aitchinson I J R 1987 {\it Acta Phys. Pol.} B {\bf 18} 207;
 reprinted in {\it Geometric Phases in Physics} edited by Shapere A 
 and Wilczek F 1989 (Singapore:  World Scientific) p. 380 
\item[] Carra P, Jerez A and Marri I 2003 \PR B {\bf 67} 045111
\item[] Cracknell A P 1975 {\it Magnetism in Crystalline Materials}
 (Oxford: Pergamon) p. 29
\item[] Dothan Y and Ne'eman Y 1966 {\it Band Spectra Generated by 
 Non-Compact Algebra} in {\it Symmetry Groups in Nuclear and Particle
 Physics - A Lecture Note and Reprint Volume} edited by Dyson F 
 (New York: Benjamin) p. 287
\item[] Edmonds A R 1974 {\it Angular Momentum in Quantum Mechanics}
 (Priceton: Princeton University) p. 107 
\item[] Fang Z Nagaosa N Takahashi K S Asamitsu A Mathieu R Ogasawara T
 Yamada H Kawasaki M Tokura Y and Terakura K 2003 {\it Science} {\bf 302} 92 
\item[] Goulon J Rogalev A Wilhelm F Goulon-Ginet C Benayoum G
 Paolasini L Brouder C Malgrange C and Metcalf P A 2000 
 \PRL {\bf 85} 4385
\item[] Goulon J, Rogalev A, Wilhelm F, Goulon-Ginet C, Carra P, Cabaret D 
 and Brouder C 2002 \PRL {\bf 88}, 237401  
\item[] Jerez A 2003 Private communication
\item[] Judd B R 1967 {\it Second Quantization and Atomic Spectroscopy}
 (Baltimore: The Johns Hopkins) p. 32 
\item[] Leinaas J M 1978 \PS {\bf 17} 483
\item[] Marri I and Carra P 2004 \PR B {\bf 69}, 0731XX (2004)
\item[] Stone M 1986 \PR D {\bf 33} 1191
\item[] Varshalovich D A, Moskalev A N and Khersonskii V K 1988 
 {\it Quantum Theory of Angular Momentum} (Singapore: World 
 Scientific) p. 202
\end{harvard}
\end{document}